\documentclass[10pt,a4paper,reqno]{amsart}
\usepackage{acronym}
\usepackage{amsfonts}
\usepackage{amsmath}
\usepackage{amssymb}
\usepackage{amsthm}
\usepackage{bm}
\usepackage{braket}
\usepackage{cite}
\usepackage{color}
\usepackage{geometry}
\usepackage{graphicx}
\usepackage{hyperref}
\usepackage{mathrsfs}
\usepackage{mathtools}
\usepackage{setspace}
\usepackage{stmaryrd}
\usepackage{subcaption}
\usepackage{tikz}

\usepackage{epsfig}
\usepackage{setspace}
\usepackage{booktabs}
\usepackage{threeparttable}
\usepackage{diagbox}

\newgeometry{top=2cm,bottom=2cm,outer=1.5cm,inner=1.5cm}
\newcommand{\bse}{\begin{subequations}}
\newcommand{\ese}{\end{subequations}}

\numberwithin{equation}{section}

\title[High-order soliton matrix for the third-order flow GI equation through the RHP ]{High-order soliton matrix  for the third-order flow equation of the Gerdjikov-Ivanov hierarchy through the Riemann-Hilbert method}

\author{JinYan Zhu}
\address[JY]{School of Mathematical Sciences, Shanghai Key Laboratory of Pure Mathematics and Mathematical Practice\\
East China Normal University \\ Shanghai 200241 \\ People's Republic of China}
\author{Yong Chen$^*$}
\address[YC]{School of Mathematical Sciences, Shanghai Key Laboratory of Pure Mathematics and Mathematical Practice \\
East China Normal University \\ Shanghai 200241 \\ People's Republic of China}
\address[YC]{College of Mathematics and Systems Science \\ Shandong University of Science and Technology \\ Qingdao 266590 \\ People's Republic of China}
\address[YC]{Department of Physics \\ Zhejiang Normal University \\ Jinhua 321004 \\ People's Republic of China}
\email{ychen@sei.ecnu.edu.cn}

\begin{document}

\begin{abstract}
The Gerdjikov-Ivanov (GI) hierarchy is derived via recursion operator, in this paper, we mainly consider the third-order flow GI equation. 
In the framework of the Riemann-Hilbert method, through a standard dressing procedure, soliton matrices for simple zeros and elementary high-order zeros in the Riemann-Hilbert problem (RHP) for the third-order flow GI equation are constructed. Taking advantage of this result, some properties and asymptotic analysis of single soliton solutions and two-soliton solutions are discussed, and the simple elastic interaction of two-soliton is proved. Compared with soliton solution of the classical second-order flow, we found that the higher-order dispersion term affects the propagation velocity, propagation direction and amplitude of the soliton. Finally, by means of certain limit technique, the high-order soliton solution matrix for the third-order flow GI equation is derived.
\end{abstract}

\maketitle

\section{Introduction}
As is well known that completely integrable equations have many important and diverse physical applications such as in water waves, plasma physics, field theory and nonlinear optics \cite{Das1989Singapore,JRS1977PRSL,KY1985JSP}. Among many integrable systems, the nonlinear Schr\"{o}dinger (NLS) equation has been recognized as a ubiquitous mathematical model, which governs weakly nonlinear and dispersive wave packets in one-dimensional physical systems. Another integrable system of NLS type, the derivative-type NLS equation
\begin{equation}\label{DNLS3}
i u_{t}+u_{x x}-i u^{2} u_{x}^{*}+\frac{1}{2} u^{3} u^{* 2}=0,
\end{equation}
where $u^{*}$ denotes the complex conjugation of $u$. Eq.\eqref{DNLS3}is first found by Gerdjikov and Ivanov in Ref.\cite{GI1983BJP} also known as the GI equation. It can be regarded as an extension of the NLS when certain higher-order nonlinear effects are taken into account, which is also known as DNLS III. In fact, there are three famous DNLS equations, another two kinds of derivative-type NLS equations are the famous Kaup-Newell (KN) equation \cite{Kaup1978JMP}
\begin{equation}\label{DNLS1}
i u_{t}+u_{x x}+i\left(u^{2} u^{*}\right)_{x}=0,
\end{equation}
which is a canonical dispersive equation derived from the Magneto-hydrodynamic equations in the
presence of the Hall effect and usually called DNLS I, and the  Chen-Lee-Liu (C-L-L) equation \cite{CLL1979PS}
 \begin{equation}
i u_{t}+u_{x x}+i u u^{*} u_{x}=0,\label{DNLS2}
\end{equation}
which appears in optical models of ultrashort pulses and is also referred to as the DNLS II. The unified expression of KN, CLL and GI equations was presented in Ref.\cite{EG2001JPA}.

In plasma physics, the GI equation \eqref{DNLS3} is a model for Alfv$\acute{e}$n waves propagating parallel to the ambient magnetic field, where $u$ being the transverse magnetic field perturbation and $x$ and $t$ being space and time coordinates, respectively \cite{EM1989PS,AGP2007AP}. In recent years, there has been much work on the GI equation, such as its Darboux transformation
and Hamiltonian structures\cite{Fan2000JMP,Fan2000JPA} the algebra-geometric solutions \cite{EG2004CSF}, the rogue wave and
breather solution \cite{SXH2012JMP}. Besides, the long-time asymptotic behavior of solution to the GI equation \eqref{DNLS3} was established in Ref.\cite{TSZT2018PAMS,XJ2013MPAG}. With the development of research, the importance of the higher-order nonlinear effects in plasma physics and other fields motivates us to consider an integrable model that possesses third dispersion and quintic nonlinearity.

In this work, we mainly consider the soliton solutions and high-order solutions of the third-order flow GI equation
\begin{equation}\label{hDNLS3}
u_t=-\frac{1}{2}u_{xxx}+\frac{3}{2}iuu_{x}u_{x}^{*}-\frac{3}{4}|u|^4u_{x}
\end{equation}
 with the help of Riemann-Hilbert method. It has been proved in Ref.\cite{EG2001JPA} that Eq.(\ref{hDNLS3}) is Liouville integrable and have multiple Hamiltonian structures.  We all known that the inverse scattering method\cite{CS1967PRL,MA1974MP,PB1984PAM} is a powerful method to solve the cauchy problem of nonlinear integrable partial differential equation, it was originally solved by using the Gel'Fand-Levitan-Marchenko (GLM) integral equation, although GLM equation can be used to obtain the solution of the equation, the solution process is very complex. Afterwards Shabat used RHP to reconstruct the inverse scattering method \cite{AS1976MP}.  As a new version of inverse scattering transform method, the Riemann-Hilbert (RH) approach has become the preferred research technique to the researchers in investigating the soliton solutions and the long-time asymptotics of integrable systems in recent years \cite{YJK2010SIAM,PD1993AM,HBB2020AR,ZN2018AM,MWX2019NARWA}.

Being an important kind of exact solution of the NLS-type equation, the high-order soliton has wide applications, it can describe a weak bound state of solitons and may appear in the study of train propagation of solitons with nearly equal velocities and amplitudes but having a particular chirp\cite{LG1994OL}, so it is necessary to study the high-order solitons of DNLS equation.

In this article, based on the recursion operator construct the GI hierarchy. In the framework of the RHP, through a standard dressing procedure, soliton matrices for simple zeros and elementary high-order zeros in the RHP for  the third-order flow GI equation are constructed, respectively. It is noted that pairs of zeros are simultaneously tackled in the situation, which is different from other NLS-type equation. Based on the determinant solution, some properties and asymptotic analysis of single soliton solution and double soliton solution are studied. Compared with the classical second-order flow GI equation, it is found that the higher-order dispersion term has a great influence on the direction, velocity and amplitude of solitons. In the case of elementary higher-order zeros, the higher-order soliton matrix of the third-order flow GI equation is derived by using the limit process of spectral parameters.

The article is arranged as follows. In Section 2, we derive the GI hierarchy with recursion operator. The RHP based on the Jost solutions to the Lax pair of the third-order flow GI equation and scattering data are constructed  in section 3. In Section 4, we discuss solutions to the regular and non-regular RHP by applying Plemelj formula. In Section 5, the N-soliton formula for the third-order flow GI equation is derived by considering the simple zeros in the RHP. In Section 6, the high-order soliton matrix is constructed, which corresponds to the elementary high-order zeros in the RHP.  The conclusion and discussion are given in the final section.

\section{Recursion Operator and the GI  Hierarchy}

In the theory of integrable equations, an important task is to construct new equations which are solvable through the inverse scattering transform method. In this section, we will associate with recursion operator, and construct the  GI hierarchy of integrable equations. The GI hierarchy has the following spectral problem:
\begin{equation}\label{y1}
Y_{x}=MY, ~~~~~~M=\left(\begin{array}{cc}
-i \lambda^2-\frac{i}{2}uv & \lambda u \\
\lambda v & i \lambda^2+\frac{i}{2}uv
\end{array}\right),\end{equation}
\begin{equation}\label{y2}
Y_{t}=NY,~~~~~~N=\left(\begin{array}{cc}
A & B\\
C & -A
\end{array}\right),\end{equation}
where $\lambda$ is the spectral parameter, $u = u(x,t)$ and $v = v(x,t)$ are field variables, $A,B$ and $C$ are the quantities depending on field variables and their derivatives and $\lambda$.

\noindent \textbf{Theorem 1}  \emph{
According to the consistency of space part \eqref{y1} and time part \eqref{y2} of spectral problem, infinite hierarchy of GI system can be obtained by recursive operator:
\begin{equation}\label{2.11}
\left(\begin{array}{c} u \\v
\end{array}\right)_{t}=(-\frac{1}{2})^{n-2}(i\sigma_3)^{n-1}(L_1\partial_x+L_2)^{n-1}\left(\begin{array}{c}
u_x \\v_x
\end{array}\right), \quad n=2,3, \ldots,
\end{equation}
where
$$
L_1=\left(\begin{array}{cc}
-1+iu\partial^{-1}v & iu\partial^{-1} u \\
-iv\partial^{-1}v& -1-iv\partial^{-1} u
\end{array}\right),~~~~~
\sigma_3=\left(\begin{array}{cc}
1 & 0 \\
0 & -1
\end{array}\right),
$$
and
$$
L_2=\left(\begin{array}{cc}
-iuv-u\partial^{-1}uv^2& u\partial^{-1}u^2v\\
v\partial^{-1}uv^2& iuv-v\partial^{-1}u^2v
\end{array}\right).
$$
}
\begin{proof}
 The corresponding zero-curvature equation or the compatibility condition of \eqref{y1} and \eqref{y2},
\begin{equation}\label{2.3}
M_{t}-N_{x}+[M,N]=0,
\end{equation}
which can get
\begin{equation}\label{2.4}
\frac{i}{2}(uv)_{t}+A_{x}-\lambda uC+\lambda vB =0,
\end{equation}
\begin{equation}\label{2.5}
\lambda u_{t}-B_{x}-2 i \lambda^2 B-iuvB-2\lambda u A =0,
\end{equation}
\begin{equation}\label{2.6}
\lambda v_{t}-C_{x}+2 i \lambda^2 C+iuvC+2\lambda v A=0 .
\end{equation}

From those equations, we can get
\begin{equation}\label{2.8}
A=\frac{-i}{2\lambda}\partial^{-1}(vB_{x}+uC_{x}+iuv^2B-iu^2vC)+A_0,
\end{equation}
where $\partial^{-1}$ is an antiderivative in $x$ which can be taken as either $\partial^{-1}=\int_{-\infty}^{x} d y$ or $\partial^{-1}=-\int_{x}^{+\infty} dy$, and $A_0$ are $x-$independent.

Using \eqref{2.8}, Eqs. \eqref{2.5} and \eqref{2.6} may be rewritten as
\begin{equation}\label{2}
 \lambda \left(\begin{array}{l}u \\v\end{array}\right)_{t}+L_1\left(\begin{array}{l}B_x \\C_x\end{array}\right)-2i\lambda^2\left(\begin{array}{l}B \\-C\end{array}\right)
+L_2 \left(\begin{array}{l}B \\C\end{array}\right)-2\lambda A_0\left(\begin{array}{l}u \\-v\end{array}\right)=0,
\end{equation}
where
$$
L_1=\left(\begin{array}{cc}
-1+iu\partial^{-1}v & iu\partial^{-1} u \\
-iv\partial^{-1}v& -1-iv\partial^{-1} u
\end{array}\right),~~~~~
L_2=\left(\begin{array}{cc}
-iuv-u\partial^{-1}uv^2& u\partial^{-1}u^2v\\
v\partial^{-1}uv^2& iuv-v\partial^{-1}u^2v
\end{array}\right).
$$

To obtain the evolution equations, we expand
\begin{equation}\label{2.10}
\left(\begin{array}{l}
B \\
C
\end{array}\right)=\sum_{j=1}^{n}\left(\begin{array}{l}
b_{j} \\
c_{j}
\end{array}\right)(\lambda)^{2j-1},
\end{equation}
Let $A_0=-2i\lambda^{2n}$. Inserting \eqref{2.10} into \eqref{2} and equating terms of the same power in $\lambda$, then we
get the following equations:
\begin{equation}
   \left(\begin{array}{l}u \\v\end{array}\right)_t+L_1\left(\begin{array}{l}b_1 \\c_1\end{array}\right)_x+L_2\left(\begin{array}{l}b_1 \\c_1\end{array}\right)=0,\label{u11}
\end{equation}
and
\begin{equation}\label{u2}
\begin{split}
 &\left(\begin{array}{l}b_{n} \\c_{n}\end{array}\right)=2\left(\begin{array}{l}u \\v\end{array}\right),~~~L_1\left(\begin{array}{l}b_{j} \\c_{j}\end{array}\right)_x-2i\sigma_3\left(\begin{array}{l}b_{j-1} \\c_{j-1}\end{array}\right)+L_2\left(\begin{array}{l}b_{j} \\c_{j}\end{array}\right)=0,~~j=2...n.
 \end{split}
\end{equation}
Eq.\eqref{u11} and \eqref{u2} are used to iterate and derive the GI hierarchy
$$
\left(\begin{array}{c} u \\v
\end{array}\right)_{t}=(-\frac{1}{2})^{n-2}(i\sigma_3)^{n-1}(L_1\partial_x+L_2)^{n-1}\left(\begin{array}{c}
u_x \\v_x
\end{array}\right), \quad n=2,3, \ldots,
$$
 the Theorem  can eventually be proved.
\end{proof}

\noindent \textbf{Remark:}  \emph{
In the zero curvature equation of GI equation, the derivative of $M$ principal diagonal to $t$ is not zero, which leads to GI hierarchy recursive form more complex than KN hierarchy \cite{GB2012JMP} and the AKNS hierarchy \cite{YJK2010SIAM}.
}

Taking $n=2$, The first nontrivial flow in the hierarchy \eqref{2.11} is
\begin{equation}\label{gi1}
\left(\begin{array}{l}u \\v\end{array}\right)_t=\left(\begin{array}{l}iu_{xx}-u^2v_x+\frac{i}{2}u^3v^2 \\-iv_{xx}-v^2u_x-\frac{i}{2}u^2v^3\end{array}\right),
\end{equation}
which form a coupled GI system. The reduction $v=-u^{*}$ for (\ref{gi1}) yields the DNLS III equation
(\ref{DNLS3}). The second nontrivial flow in the hierarchy (\ref{2.11}) is
\begin{equation}
\left(\begin{array}{l}u \\v\end{array}\right)_t=\left(\begin{array}{l}-\frac{1}{2}u_{xxx}-\frac{3i}{2}uu_{x}v_{x}-\frac{3}{4}(u^2v^2)u_{x} \\-\frac{1}{2}v_{xxx}+\frac{3i}{2}u_{x}v_{x}v-\frac{3}{4}(u^2v^2)v_{x}\end{array}\right),\label{hc2}
\end{equation}
Taking $v=-u^{*}$, the Eq.(\ref{hc2})is simplified as
$$
u_t=-\frac{1}{2}u_{xxx}+\frac{3}{2}iuu_{x}u_{x}^{*}-\frac{3}{4}|u|^4u_{x},
$$
which corresponding to Eq.(\ref{hDNLS3}). The explicit forms for $b_j$ and $c_j$ in the present case are
\begin{equation}
\begin{split}
&b_3=2u, ~c_3=-2u^{*}, ~b_2=iu_{x}, ~c_2=iu^{*}_{x},\\
&b_1=-\frac{1}{2}u_{xx}+\frac{1}{2}iu^2u^{*}_{x}-\frac{1}{4}u^3u^{*2},\\
&c_1=\frac{1}{2}u^{*}_{xx}+\frac{1}{2}iu^{*2}u_{x}+\frac{1}{4}u^{*3}u^2.
\end{split}
\end{equation}

For convenience, we list down the explicit spectral problem of Eq.(\ref{hDNLS3}) here
\begin{align}\label{yx}
Y_{x}=MY,~~M=-i\lambda^2 \sigma_3+\lambda Q-\frac{i}{2}Q^2\sigma_3,
\end{align}
\begin{align}\label{yt}
Y_{t}=NY,~~N=-2i\lambda^6 \sigma_3+Z_5\lambda^{5}+Z_4\lambda^{4}+Z_3\lambda^3+Z_2\lambda^2+Z_1\lambda+Z_0,
\end{align}
where
\begin{equation}\label{ytt}
\begin{split}
&Z_5=2 Q,\quad Z_4=-i Q^{2} \sigma_{3},\quad Z_3=i \sigma_{3}Q_{x},\\
&Z_2= -\frac{1}{2}[Q,Q_{x}]+\frac{1}{4} i Q^{4} \sigma_{3},\\
&Z_1=-\frac{1}{2} Q_{x x}+\frac{ i}{2}  Q Q_{x}Q \sigma_{3}-\frac{1}{4} Q^{5},\\
&Z_0=\frac{i}{4}(QQ_{xx}+Q_{xx}Q)\sigma_{3}-\frac{i}{4}Q_{x}^2\sigma_{3}+\frac{i}{8}Q^6\sigma_{3}.
\end{split}\end{equation}
\begin{equation}\label{Q}
Q=\left(\begin{array}{cc}
0 & u \\
-u^{*} & 0
\end{array}\right),
\end{equation}
it's easy to see that
$$Q^{\dagger} = -Q, ~~~~~~~\sigma_{3} Q \sigma_{3}=-Q,$$
where the superscript $'\dagger'$ represents the Hermitian of a matrix, and $[A, B]$ denotes AB-BA.

\section{The construction of RHP}

This section mainly constructs the RHP of Eq. (\ref{hDNLS3}). In the following analysis, we treat $Y$ in Eqs.(\ref{yx}) and (\ref{yt}) as a fundamental matrix of those linear equations. In our analysis, we mainly consider the zero boundary condition, i.e.
$$
u(x,t_0)\rightarrow0,~~~x\rightarrow\pm\infty,
$$
which belongs to Schwartz space. Therefore, it is easy to take the form of the solution of Eqs.(\ref{yx}) and (\ref{yt}) as
\begin{equation}\label{yj}
Y=J e^{(-i \lambda^2 x-2 i \lambda^{6} t)\sigma_{3}}.
\end{equation}
The Lax pair of Eq.(\ref{yx})-(\ref{yt}) becomes
\begin{equation}\label{jx}
J_{x}+\mathrm{i} \lambda^{2}\left[\sigma_{3},J\right]=(\lambda Q -\frac{i}{2}Q^2\sigma_3)J,
\end{equation}
\begin{equation}\label{jt}
J_{t}+2i\lambda^{6}\left[\sigma_{3}, J\right]=(Z_5\lambda^{5}+Z_4\lambda^{4}+Z_3\lambda^3+Z_2\lambda^2+Z_1\lambda+Z_0) J.
\end{equation}
where $Q,Z_i(i=0...5)$ has been given by  Eq.(\ref{ytt}),(\ref{Q}).

In this consideration, the time $t$ is fixed and is a dummy variable, and thus it will be
suppressed in our notation. In the scattering problem, we first introduce matrix Jost solutions $J(x,\lambda)$ of Eq.(\ref{jx})
with the following asymptotics at large distances
 \begin{equation}\label{JI}
 J(x,\lambda)\rightarrow I,~~~~~~~x\rightarrow\pm\infty.
 \end{equation}
It is easy to find that $J(x,\lambda)$ satisfies the following integral equation
\begin{equation}\label{jm}
J_{-}(x,\lambda)=I+ \int_{-\infty}^{x} e^{\mathrm{i} \lambda^{2} \sigma_{3}(y-x)} (\lambda Q(y)-\frac{i}{2}Q^2\sigma_3) J_{-} e^{\mathrm{i} \lambda^{2} \sigma_{3}(x-y)} \mathrm{d} y,
\end{equation}
\begin{equation}\label{jp}
J_{+}(x,\lambda)=I-\int_{x}^{+\infty} e^{\mathrm{i} \lambda^{2} \sigma_{3}(y-x)}(\lambda Q(y)-\frac{i}{2}Q^2\sigma_3) J_{+} e^{\mathrm{i} \lambda^{2} \sigma_{3}(x-y)} \mathrm{d} y.
\end{equation}
From the above Volterra type integral equations, we can easily prove the existence and uniqueness of the Jost solutions through standard iteration method.  Partitioning $J_{\pm}$ into columns as $J=(J^{[1]}, J^{[2]})$, due to the structure Eq.(\ref{jm}) of the potential $Q$, we have

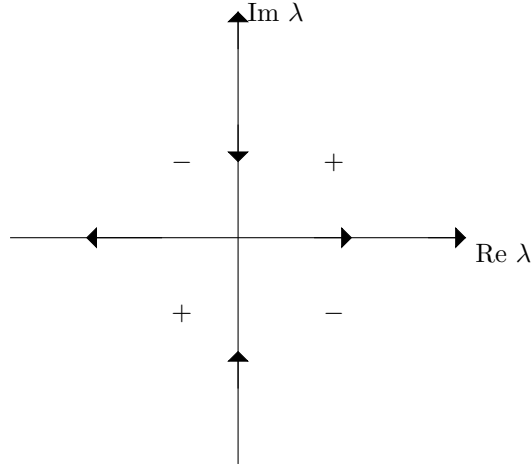
\begin{figure}
\center
\begin{tikzpicture}\usetikzlibrary{arrows}
\coordinate [label=0: Im $\lambda$] ()at (0,3);
\coordinate [label=0: Re $\lambda$] ()at (3,-0.2);
\coordinate [label=0:] ()at (2,0.1);
\coordinate [label=0:] ()at (-2.6,0.1);
\coordinate [label=0: $+$] ()at (1,1);
\coordinate [label=0: $-$] ()at (-1,1);
\coordinate [label=0: $+$] ()at (-1,-1);
\coordinate [label=0: $-$] ()at (1,-1);
\draw[->, >=triangle 90,] (1,0)--(1.5,0);
\draw[->, >=triangle 90,] (2.5,0)--(3,0);
\draw[->, >=triangle 90,] (-1,0)--(-2,0);
\draw[->, >=triangle 90,] (0,1.5)--(0,1);
\draw[->, >=triangle 90,] (0,2.5)--(0,3);
\draw[->, >=triangle 90,] (0,-2)--(0,-1.5);
\draw[thin](0,3)--(0,-3);
\draw[thin](3,0)--(-3,0);
\end{tikzpicture}
\caption{\small The jump contour in the complex $\lambda$-plane. The positive (negative) side lies on the left (right) as one traverses the
contour. }
\label{hdnls-fig.1}
\end{figure}

\noindent \textbf{Proposition 3.1}  \emph{
The column vectors $J_{-}^{(1)}$ and $J_{+}^{(2)}$ are continuous for $\lambda \in D_{+} \cup \mathbb{R} \cup i \mathbb{R}$ and analytic for $\lambda \in D_{+}$, while the columns $J_{+}^{(1)}$and $J_{-}^{(2)}$ are continuous for $\lambda \in D_{-} \cup \mathbb{R} \cup i\mathbb{R}$ and analytical for $\lambda \in D_{-}$,
where
$$
D_{+}=\left\{\lambda \mid \arg \lambda \in\left(0, \frac{\pi}{2}\right) \cup\left(\pi, \frac{3 \pi}{2}\right)\right\}, \quad D_{-}=\left\{\lambda \mid \arg \lambda \in\left(\frac{\pi}{2}, \pi\right) \cup\left(\frac{3 \pi}{2}, 2 \pi\right)\right\}.
$$
}
The distribution area of D is shown in Fig.(\ref{hdnls-fig.1}).

In fact, the $J_{+} E$ and $J_{-} E$ are the simultaneous solutions for the Lax pair \eqref{jx}. Therefore, they have following linear relation by  the constant scattering matrix $S(\lambda)$
\begin{equation}\label{jmp}
J_{-} E=J_{+} E S(\lambda), \quad \lambda \in \mathbb{R} \cup \mathrm{i} \mathbb{R},
\end{equation}
where $E=e^ {-\mathrm{i} \lambda^{2} x \sigma_{3}}$ and $S(\lambda)=\left(s_{i j}\right)_{2 \times 2}$.
By using Abel's  formula and $\operatorname{tr}(Q) = 0$, we obtain that the determinant of $J$ is independent of $x$,
then considering the boundary conditions (\ref{JI}), we can get
 $$
\operatorname{det} J=1.
$$
Thus we can derive $\operatorname{det} S(\lambda)=1$.

\noindent \textbf{Proposition 3.2}  \emph{
Through the analytic property of $J_{-}$, it's easy to know that $s_{11}$ allows analytic extension to $D_{+}$, $s_{22}$ can be analytically extended to $D_{-}$.
}

\begin{proof}
 According to the relation (\ref{jmp}) we have
\begin{equation}\label{s}
S(\lambda)=\lim _{x \rightarrow+\infty} E^{-1} J_{-} E=I+ \int_{-\infty}^{+\infty} E^{-1}(\lambda Q(y)-\frac{i}{2}Q^2\sigma_3) J_{-} E \mathrm{d} x, \quad \lambda \in \mathbb{R} \cup \mathrm{i} \mathbb{R}.
\end{equation}
So
\begin{equation}\label{2.51}
S(\lambda)=I+\lambda\left(\begin{array}{cc}
\int_{-\infty}^{+\infty}uJ_{-}^{21} d x & \int_{-\infty}^{+\infty}uJ_{-}^{22}e^{2i\lambda^2x} d x \\
-\int_{-\infty}^{+\infty}u^{*}J_{-}^{11}e^{-2i\lambda^2x} d x & -\int_{-\infty}^{+\infty}u^{*}J_{-}^{12} d x
\end{array}\right),\end{equation}
i,e
$$
s_{11}=1+\lambda\int_{-\infty}^{+\infty}uJ_{-}^{21} d x,~~s_{22}=1-\lambda\int_{-\infty}^{+\infty}u^{*}J_{-}^{12} d x,
$$
through the analytic property of $J_{-}$, it's easy to know that $s_{11}$ allows analytic extension to $D_{+}$, $s_{22}$ can be analytically extended to $D_{-}$.
\end{proof}

In order to construct the RHP, introducing the notation
 \begin{equation}
P_{+}=(J_{-}^{[1]}, J_{+}^{[2]})=J_{-} H_{1}+J_{+} H_{2}=J_{+} E\left(\begin{array}{cc}
s_{11}&0 \\
s_{21}& 1\\
\end{array}\right) E^{-1},
\end{equation}
where $H_{1}=\operatorname{diag}\{1,0\}$ and $H_{2}=\operatorname{diag}\{0,1\}$.
Through the previous analysis, we can see that $P_{+}$ is analytic in  $D_{+}$ and $\operatorname{det}\left(P_{+}\right)=s_{11}.$ To find the boundary condition of $P_{+}$ as $\lambda \rightarrow \infty,$ we consider the following asymptotic expansion
\begin{equation}\label{expandp}
P_{+}=P_{+}^{(0)}+\frac{1}{\lambda} P_{+}^{(1)}+\frac{1}{\lambda^{2}} P_{+}^{(2)}+O\left(\frac{1}{\lambda^{3}}\right).
\end{equation}
Substituting (\ref{expandp}) into (\ref{jx}) and equating terms with like powers of $\lambda$, which lead to
\begin{equation}\label{2.24}
P_{+x}^{(0)}=0,
\end{equation}
without loss of generality, we can set $ P_{+}^{(0)}= I$. This means
 \begin{equation}\label{bj}
 P_{+}\rightarrow I,~~~~~\lambda\in D_{+}\rightarrow\infty.
 \end{equation}.

 To obtain the analytic counterpart of $P_+$ in $D_{-}$, we consider the adjoint scattering equation of (\ref{jx})
\begin{equation}\label{2.25}
\Phi_{x}=-i\lambda^2\left[\sigma_{3}, \Phi\right]-\lambda\Phi Q+\frac{i}{2}\Phi Q^2\sigma_3,
\end{equation}
it is easy to see that $J^{-1}$ is the solution of the adjoint Eq.(\ref{2.25}) and satisfy the boundary condition $J^{-1} \rightarrow I$ as $x \rightarrow \pm \infty.$  Taking the similar procedure as above denote matrices $J^{-1}$ as a collection of rows
\begin{equation}
J_{+}^{-1}=\left((J_{+}^{-1})^{[1]},(J_{+}^{-1})^{[2]}\right)^{T}, \quad J_{-}^{-1}=\left((J_{-}^{-1})^{[1]},(J_{-}^{-1})^{[2]}\right)^{T},
\end{equation}
we can show that the adjoint Jost solutions
\begin{equation}P_{-}^{-1}=H_{1} J_{-}^{-1}+H_{2} J_{+}^{-1}= E\left(\begin{array}{cc}
\hat{s}_{11}&\hat{s}_{12} \\
0& 1\\
\end{array}\right) E^{-1}J_{+}^{-1}\end{equation}
analytic for $D_{-}$, where
$$
J_{-}^{-1}=E S^{-1} E^{-1} J_{+}^{-1},~~~~~~
\hat{S}=\left(\begin{array}{cc}
\hat{s}_{11}&\hat{s}_{12} \\
\hat{s}_{21}&\hat{s}_{22}\\
\end{array}\right),
$$
and $\operatorname{det} P_{-}^{-1}=\hat{s}_{11}.$ Through direct calculation, we can get that $P_{-}^{-1}$ also satisfies the same boundary condition (\ref{bj}),  as $\lambda\rightarrow\infty$. i.e.
\begin{equation}\label{bl}
P^{-1}_{-}(x, \lambda) \rightarrow I, \quad \lambda \in \mathbb{C}_{-} \rightarrow \infty.
\end{equation}
Hence,  we have constructed two matrix functions $P_{\pm}(x,\lambda)$ which are analytic for $\lambda\in D_{\pm}$, respectively. Thus the RHP  can be constructed as follow by $P_{+},P^{-1}_{-}$
\begin{equation}\label{2.26}
P^{-1}_{-}(x, \lambda) P_{+}(x, \lambda)=G(x, \lambda)=E\left(\begin{array}{cc}
1 & \hat{s}_{12}\\
s_{21} & 1  \\
\end{array}\right) E^{-1}, \quad \lambda \in \mathbb{R} \cup \mathrm{i} \mathbb{R},
\end{equation}
with boundary condition
\begin{equation}\label{L5}
P_{\pm} \rightarrow I \quad  , \quad \lambda \rightarrow \infty.
\end{equation}

At the end of this section, we consider the time evolution of the scattering matrices $S(\lambda)$ and $\hat{S}(\lambda)$, since $J$ satisfies the temporal Eq.(\ref{jt}) of the Lax pair and the relation (\ref{jmp}), then according to the evolution property (\ref{jmp}) and $Q \rightarrow 0, Z_{i}(i=1...5)\rightarrow 0$ as $|x| \rightarrow \infty,$ we have
$$
S_{t}+2 \mathrm{i} \lambda^{6}\left[\sigma_{3}, S\right]=0.
$$
And then the time evolution of $\hat{S}(\lambda)$ can be gotten immediately
$$
\hat{S}_{t}+2 \mathrm{i} \lambda^{6}[\sigma_{3}, \hat{S}]=0.
$$
These two equations lead that
$$
s_{11, t}=\hat{s}_{11, t}=0,
$$
\begin{equation}\label{L16}
s_{12}(t ; \lambda)=s_{12}(0 ; \lambda) \exp \left(-4 \mathrm{i} \lambda^{6} t\right), \quad \hat{s}_{21}(t ; \lambda)=\hat{s}_{21}(0 ; \lambda) \exp \left(4 \mathrm{i} \lambda^{6} t\right).
\end{equation}

\section{Solution of the RHP}

In this section, we discuss how to solve the matrix RHP (\ref{2.26}) in the complex $\lambda$ plane.
The RHP (\ref{2.26}) constructed in above section is regular when $\operatorname{det}\left(P_{+}\right)=s_{11}$ $\neq 0$ and $\operatorname{det}(P_{-}^{-1})=\hat{s}_{11} \neq 0$ for all $\lambda$, and is nonregular when $\operatorname{det}(P_{+})$ and $\operatorname{det}(P_{-})$ can be zero at certain discrete locations of $\lambda$. In fact, a non-regular RHP can be transformed into a regular one, thus we consider the regular case at first.

\subsection{Solution to the Regular RHP}\

In this subsection, we first consider the regular RHP of (\ref{2.26}), i.e.,  in their analytic domain. Rewriting Eq. (\ref{2.26}) as
\begin{equation}
\left(P^{+}\right)^{-1}(\lambda)-P^{-1}_{-}(\lambda)=\widehat{G}(\lambda)\left(P^{+}\right)^{-1}(\lambda), \quad \lambda \in \mathbb{R} \cup \mathrm{i} \mathbb{R},
\end{equation}
where
\begin{equation}\label{2.36}
\widehat{G}=I-G=-E\left(\begin{array}{ccc}
0 & \hat{s}_{12} \\
s_{21} & 0\\
\end{array}\right) E^{-1}.\end{equation}
By Plemelj formula, the formal solution of this problem reads as
\begin{equation}\label{2.35}
\left(P^{+}\right)^{-1}(\lambda)=I+\frac{1}{2 \pi i} \int_{T} \frac{\widehat{G}(\xi)\left(P^{+}\right)^{-1}(\xi)}{\xi-\lambda} d \xi, \quad \lambda \in D_{+},
\end{equation}
and $T=(-\mathrm{i} \infty, 0]\cup(\mathrm{i} \infty, 0]\cup[0,-\infty)\cup[0, \infty) $.

Under the canonical normalization condition (\ref{L5}), the solution to this regular RHP is unique. Suppose (\ref{2.26}) has two sets of solutions $P_{\pm}$ and $\tilde{P}_{\pm}$. Then
$$
P_{-}^{-1}(\lambda) P_{+}(\lambda)=\tilde{P}_{-}^{-1}(\lambda) \tilde{P}_{+}(\lambda),
$$
and thus
\begin{equation}
\tilde{P}_{-}(\lambda) P_{-}^{-1}(\lambda)=\tilde{P}_{+}(\lambda) P_{+}^{-1}(\lambda), \quad \lambda \in \mathbb{R} \cup \mathrm{i} \mathbb{R}.
\end{equation}
Since $\tilde{P}_{-}(\lambda) P_{-}^{-1}(\lambda)$ and $\tilde{P}_{+}(\lambda) P_{+}^{-1}(\lambda)$ are analytic in $D_{-}$ and $D_{+}$ respectively, and they are equal to each other on $\mathbb{R} \cup \mathrm{i} \mathbb{R}$, they together define a matrix function which is analytic in the whole plane of $\lambda$. Due to the boundary condition (\ref{L5}), we have
\begin{equation}
\tilde{P}_{-}^{-1}(\lambda) P_{-}(\lambda)=\tilde{P}_{+}(\lambda) P_{+}^{-1}(\lambda)=I,
\end{equation}
for all $\lambda$ by applying the Liouville's theorem. That is, $\tilde{P}_{\pm}=P_{\pm}$, which implies the uniqueness of solution to the above RHP (\ref{2.26}).

\subsection{Solution to the Nonregular RHP}
\

In the more general case, the RHP (\ref{2.26}) is not regular, i.e., $\operatorname{det} P_{+}(\lambda)=s_{11}(\lambda)$ and $\operatorname{det} P_{-}^{-1}(\lambda)=\hat{s}_{11}(\lambda)$ can be zero at certain discrete locations. In order to study a nonregular RHP, we shall consider symmetric property of these zero points. Note that $s_{11}(\lambda)$ and $\hat{s}_{11}(\lambda)$ are time independent, so the roots of $s_{11}(\lambda)$ and $\hat{s}_{11}(\lambda)$ are also time independent.

The Hermitian of the spectral equation (\ref{jx}) reads as
\begin{equation}\label{2.27}
\left(J^{\dagger}\right)_{x}=-\mathrm{i} \lambda^{2}\left[\sigma_{3}, J^{\dagger}\right]- \lambda J^{\dagger} Q+\frac{i}{2}J^{\dagger} Q^2\sigma_3,
\end{equation}
where $Q^{\dagger} = -Q$ is used. Thus $J^{\dagger}(x,\lambda^{*})$ satisfies the adjoint scattering Eq.(\ref{2.25}). $J^{\dagger}\left(x, \lambda^{*}\right)$ and $J^{-1}(x, \lambda)$ must be linearly dependent on each other. Recalling the boundary conditions of Jost solutions $J,$ we further see that $J^{\dagger}\left(x,\lambda^{*}\right)$ and $J^{-1}(x, \lambda)$ have the same boundary conditions at $x \rightarrow \pm \infty$ and hence they must be the same solutions of the adjoint Eq.(\ref{2.25}) i.e.
\begin{equation}\label{2.28}
J^{\dagger}\left(x, \lambda^{*}\right)=J^{-1}(x, \lambda),
\end{equation}
so there is
\begin{equation}\label{2.29}
\left(P_{+}\right)^{\dagger}\left(\lambda^{*}\right)=P^{-1}_{-}(\lambda).\end{equation}
In addition, in view of the scattering relation (\ref{jmp}) between $J_{+}$ and $J_{-}$,
it's easy to know that $S(\lambda)$ also satisfies involution property
\begin{equation}\label{2.30}
S^{\dagger}\left(\lambda^{*}\right)=S^{-1}(\lambda).
\end{equation}

 Besides, from the symmetric property $\sigma_{3} Q \sigma_{3}=-Q$ and $\sigma_{3} Q^2\sigma_{3}=Q$, we conclude that
 \begin{equation}\label{2.31}
J(\lambda)=\sigma_{3} J(-\lambda) \sigma_{3}.
\end{equation}
It follows that
\begin{equation}\label{2.32}
P_{\pm}(-\lambda)=\sigma_{3} P_{\pm}(\lambda) \sigma_{3},
\end{equation}
and
\begin{equation}\label{2.33}
S(-\lambda)=\sigma_{3} S(\lambda) \sigma_{3}.
\end{equation}

From the (\ref{2.30}) and (\ref{2.33}), we obtain the relations
\begin{equation}\label{2.34}
s_{11}^{*}\left(\lambda^{*}\right)=\hat{s}_{11}(\lambda),  s_{21}^{*}\left(\lambda^{*}\right)=\hat{s}_{12}(\lambda),  s_{12}^{*}\left(\lambda^{*}\right)=\hat{s}_{21}(\lambda), ~~~\lambda \in \mathbb{R} \cup i \mathbb{R},
\end{equation}
and
\begin{equation}\label{2.37}
s_{1 1}(\lambda)=s_{1 1}(-\lambda), s_{2 2}(\lambda)=s_{2 2}(-\lambda),
s_{1 2}(-\lambda)=-s_{1 2}(\lambda), s_{2 1}(-\lambda)=-s_{2 1}(\lambda).
\end{equation}
Thus $s_{11}(\lambda)$ is an even function, and each zero $\lambda_{k}$ of $s_{11}$ is accompanied with zero $-\lambda_{k}$. Similarly, $\hat{s}_{11}(\lambda)$ has two zeros $\pm\bar{\lambda}_{k}$.

Here we first consider the case of simple zeros $\{\pm\lambda_{k} \in D_{+},1 \leq k \leq N\}$ and $\{\pm\bar{\lambda}_{k} \in D_{-}, 1 \leq k \leq N\},$ where $N$ is the number of these zeros. Due to the involution property (\ref{2.34}), the involution relation is obtained as follow
\begin{equation}\label{2.38}
\bar{\lambda}_{k}=\lambda_{k}^{*}.
\end{equation}
It follows that symmetry relation (\ref{2.34})and (\ref{2.37}), in this case, both $\operatorname{ker}\left(P_{+}\left(\pm \lambda_{k}\right)\right)$ and $\operatorname{ker}(P_{-}^{-1}\left(\pm \bar{\lambda}_{k}\right))$ are one-dimensional and spanned by single column vector $\left|v_{k}\right\rangle$ and single row vector $\left\langle v_{k}\right|,$ respectively, thus
\begin{equation}\label{2.40}
P_{+}\left(\lambda_{k}\right) \left|v_{k}\right\rangle=0, \quad \left\langle v_{k}\right| P_{-}^{-1}\left(\bar{\lambda}_{k}\right)=0, \quad 1 \leq k \leq N.
\end{equation}
By the symmetry relation (\ref{2.29}), it is easy to get
\begin{equation}\label{2.39}
\left|v_{k}\right\rangle=\left\langle v_{k}\right|^{\dagger}.
\end{equation}

Differentiating both sides of the first equation of (\ref{2.40}) with respect to $x$ and $t,$ and recalling the Lax (\ref{jx})-(\ref{jt}) we have
$$
P_{+}(\lambda_{k} ; x)\left(\frac{d|v_{k}\rangle}{d x}+\mathrm{i} \lambda^{2} \sigma_{3}|v_{k}\rangle\right)=0, \quad P_{+}(\lambda_{k} ; x)\left(\frac{d|v_{k}\rangle}{d t}+2 \mathrm{i} \lambda^{6} \sigma_{3}|v_{k}\rangle\right)=0.
$$
It concludes that
$$
\left|v_{k}\right\rangle=e^{-\mathrm{i} \lambda_{k}^{2} \sigma_{3} x-2 \mathrm{i} \lambda_{k}^{6} \sigma_{3} t}\left|v_{k0}\right\rangle\mathrm{e}^{\int_{x_{0}}^{x} \alpha_{k}(y) \mathrm{d} y+\int_{t_{0}}^{t} \beta_{k}(\tau) \mathrm{d} \tau},
$$
where $v_{k 0}=\left.v_{k}\right|_{x=0}$ and $\alpha_{k}(x)$ and $\beta_{k}(t)$ are two scalar functions.

Based on above analysis, we have the following theorem for the solution to the nonregular RHP with canonical normalization condition (\ref{L5}).

\noindent \textbf{Theorem 2}  \emph{
The solution to a nonregular RHP (\ref{2.26}) with simple zeros under
the canonical normalized condition (\ref{bj}) and (\ref{bl}) is
\begin{equation}\label{2.45}
P_{+}=\hat{P}_{+} \Gamma,~~~~P_{-}^{-1}=\Gamma^{-1}\hat{P}_{-}^{-1},
\end{equation}
where
$$
\Gamma(\lambda)=\Gamma_{N}(\lambda) \Gamma_{N-1}(\lambda) \cdots \Gamma_{1}(\lambda),~~~~~
\Gamma^{-1}(\lambda)=\Gamma_{1}^{-1}(\lambda) \Gamma_{2}^{-1}(\lambda) \cdots \Gamma_{N}(\lambda),
$$
\begin{equation}\label{G1}
\Gamma_{k}(\lambda)=I+\frac{A_{k}}{\lambda-\lambda_{k}^{*}}-\frac{\sigma_{3} A_{k} \sigma_{3}}{\lambda+\lambda_{k}^{*}},
\end{equation}
\begin{equation}\label{G-1}
\Gamma_{k}^{-1}(\lambda)=I+\frac{A_{k}^{\dagger}}{\lambda-\lambda_{k}}-\frac{\sigma_{3} A_{k}^{\dagger} \sigma_{3}}{\lambda+\lambda_{k}}, ~~~~~k=1,2, \ldots, N
\end{equation}
\begin{equation}
A_{k}=\frac{\lambda_{k}^{* 2}-\lambda_{k}^{2}}{2}\left(\begin{array}{cc}
\alpha_{k}^{*} & 0 \\
0 & \alpha_{k}
\end{array}\right)\left|w_{k}\right\rangle\left\langle w_{k}\right|,~~~~
\alpha_{k}^{-1}=\langle w_{k}|\left(\begin{array}{cc}
\lambda_{k} & 0 \\
0 & \lambda_{k}^{*}
\end{array}\right)| w_{k}\rangle,
\end{equation}
and
\begin{equation}\label{L10}
\operatorname{det} \Gamma_{k}(\lambda)=\frac{\lambda^{2}-\lambda_{k}^{2}}{\lambda^{2}-\lambda_k^{* 2}},
~~\left|w_{k}\right\rangle=\Gamma_{k-1}\left(\lambda_{k}\right) \cdots \Gamma_{1}\left(\lambda_{k}\right)\left|v_{k}\right\rangle, ~~\left\langle w_{k}|=| w_{k}\right\rangle^{\dagger},
\end{equation}
Therefore, $\Gamma(x,t,\lambda)$ and $\Gamma^{-1}(x,t,\lambda)$ accumulates all zero of the RHP, and then we obtain the regular RHP
\begin{equation}\label{rhp2}
\hat{P}_{-}^{-1}(\lambda) \hat{P}_{+}(\lambda)=\Gamma(\lambda) G(\lambda) \Gamma^{-1}(\lambda), \quad \lambda \in \mathbb{R} \cup i \mathbb{R},
\end{equation}
and the boundary condition
$\hat{P}_{\pm} \rightarrow I$ as $\lambda \rightarrow \infty$, where $\hat{P}_{\pm}$ are analytic in $D_{\pm}$ respectively.
}

The proof has been given in Ref. \cite{ZYS2017JMP}, and we will not repeat it here.

\section{The inverse problem}

  The ultimate purpose of inverse scattering is to obtain the potential $u$. Based on (\ref{expandp}), the potential can be obtained from the asymptotic expansion of Jost solutions $P$ as $\lambda \rightarrow+\infty$,
\begin{equation}\label{2.43}
Q=i[\sigma_3, P_{+}^{(1)}],
\end{equation}
from this formula, we can get the potential
\begin{equation}\label{u1}
u=2i(P_{+}^{(1)})_{12}.
\end{equation}

It is well known that the soliton solutions correspond to the vanishing of scattering coefficients, $G=I, \hat{G}=0$. Thus, we intend to solve the corresponding RHP(\ref{rhp2}). The product representations $\Gamma(\lambda)$ and $\Gamma^{-1}(\lambda)$ are not convenient
to use for later calculations in the inverse scattering transform method, it is necessary to simplify the expression of $\Gamma(\lambda)$ and its inverse, in fact,
\begin{equation}\label{gamma}
 \Gamma(\lambda)=I+\sum_{j=1}^{N}\left[\frac{B_{j}}{\lambda-\lambda_{j}^{*}}-\frac{\sigma_{3} B_{j} \sigma_{3}}{\lambda+\lambda_{j}^{*}}\right],
\end{equation}
and
$$
\Gamma^{-1}(\lambda)=I+\sum_{j=1}^{N}\left[\frac {B_{j}^{\dagger} }{\lambda-\lambda_{j}}-\frac{\sigma_{3}  B_{j}^{\dagger} \sigma_{3}}{\lambda+\lambda_{j}}\right],
$$
with $B_{j}=\left|z_{j}\right\rangle\left\langle v_{j}\right|$. To determine the form of matrix $B_{j}$, we consider $ \Gamma(\lambda)\Gamma(\lambda)^{-1}=I$. Taking into account the  residue condition at $\lambda_{j}$, we have
$$
\operatorname{Res}_{\lambda=\lambda_{j}} \Gamma(\lambda) \Gamma^{-1}(\lambda)= \Gamma(\lambda_j)B_{j}^{\dagger}=0,
$$
and it yields
\begin{equation}
\left[I+\sum_{k=1}^{N}\left(\frac{\left|z_{k}\right\rangle\left\langle v_{k}\right| }{\lambda_{j}-\lambda_{k}^{*}}-\frac{\sigma_{3}\left|z_{k}\right\rangle\left\langle v_{k}\right|  \sigma_{3}}{\lambda_{j}+\lambda_{k}^{*}}\right)\right]\left|v_{j}\right\rangle=0, \quad j=1,2, \ldots N
\end{equation}
it's easy to figure out
\begin{equation}
\left|z_{k}\right\rangle_{1} =\sum_{j=1}^{N} (M^{-1})_{jk}\left|v_{j}\right\rangle_{1},
\end{equation}
where $\left|z_{k}\right\rangle_{l}$ denotes the $l-$th element of $\left|z_{k}\right\rangle$, matrix $M$ is defined as
\begin{equation}\label{mjk}
(M)_{jk}=\frac{\left\langle v_{k}\left|\sigma_{3}\right| v_{j}\right\rangle}{\lambda_{j}+\lambda_{k}^{*}}-\frac{\langle v_{k}\mid v_{j}\rangle}{\lambda_{j}-\lambda_{k}^{*}}.
\end{equation}

From these equations enable us to have
$$
P_{+}^{(1)}=\sum_{j=1}^{N}(B_{j}-\sigma_{3} B_{j} \sigma_{3}),
$$
by Eq.(\ref{2.43}), we can obtain that the potential function $u$ is
\begin{equation}\label{2.48}
u=2i\left[\sum_{j=1}^{N}(B_{j}-\sigma_{3} B_{j} \sigma_{3})_{12}\right],
\end{equation}
and substituting above expressions for $\left|z_{k}\right\rangle_{l}$ and $\left|v_{j}\right\rangle_{l}$ into Eq.(\ref{2.48}) gives
\begin{equation}\label{L20}
u=-4i\frac{detF}{detM},
\end{equation}
where $M$ defined as (\ref{mjk}), and
$$
F=\left[\begin{array}{cccc}
M_{11} & \cdots & M_{1 N} & \left|v_{1}\right\rangle_{1} \\
\vdots & \ddots & \vdots & \vdots \\
M_{N 1} & \cdots & M_{N N} & \left|v_{N}\right\rangle_{1} \\
\left\langle\left. v_{1}\right|_{2}\right. & \cdots & \left\langle\left. v_{N}\right|_{2}\right. & 0
\end{array}\right].
$$
Based on the dressing method \cite{SN1984CB}, it is straightforward to verify that (\ref{L20}) satisfies the higher-order GI equation.

Next, we mainly obtain the soliton solutions of the third-order flow GI equation. To get the explicit $N$-soliton solutions, we may take
$$
\left|v_{k}\right\rangle=\left(\begin{array}{c}
c_{k}e^{\theta_{k}} \\
e ^{-\theta_{k}}\\
\end{array}\right),~~~~~\left\langle v_{k}\right|=(\begin{array}{cc}
c_{k}^{*}e^{\theta^{*}_{k}} & e ^{-\theta^{*}_{k}}
\end{array}),
$$
where $\theta_{k}=-i \lambda_{k}^2 x-2 i \lambda_{k}^{6}t$. Let  $\lambda_{j}=\xi_j+i \eta_j,$
then
$$
\begin{array}{l}
z_{j}=2 m_{j}(x-(8 m_j^{2}-6 \beta_{j}^2) t), ~~~~~\phi_{j}=- \beta_{j} x-2(\beta_{j}^{3}-12 m_{j}^{2}v_j) t, \\
m_{j}=\xi_j \eta_j, \quad \beta_{j}=\xi_j^{2}-\eta_j^{2},
\end{array}
$$
where $z_{j} ,\phi_{j}$ are the real and imaginary parts of $\theta_{j}$. In what follows, we will investigate the properties of the single-soliton and two-soliton solutions in more details.

\subsection{ Single-soliton solution}
\

To obtain the single-soliton solution, we set $N = 1$ in formula (\ref{L20}). The solution is
\begin{equation}\label{L21}
u(x, t)=-2i(\lambda_{1}^2-\lambda_{1}^{*2}) \frac{c_{1} e^{\theta_{1}-\theta_{1}^{*}}}{\lambda_{1} e^{-(\theta_{1}+\theta_{1}^{*})}+\lambda_{1}^{*}|c_{1}|^{2} e^{\theta_{1}+\theta_{1}^{*}}},
\end{equation}

\begin{figure}
\centering
\includegraphics[width=3.6cm,height=3.0cm]{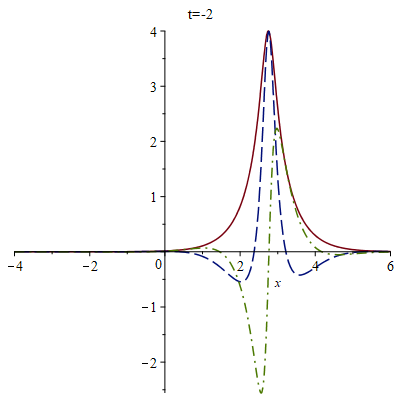}\qquad \qquad
\includegraphics[width=3.6cm,height=3.0cm]{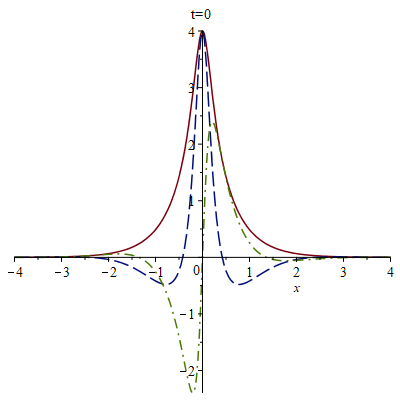}\qquad \qquad
\includegraphics[width=3.6cm,height=3.0cm]{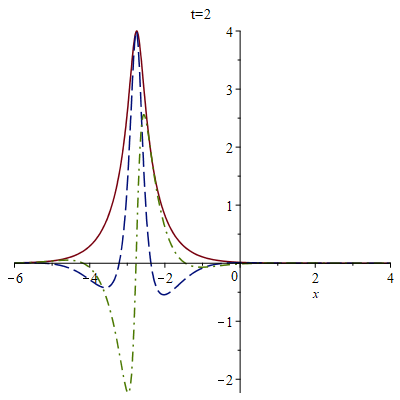}
\caption{(Color online) One-soliton $u(x,t)$ in (\ref{L21}) with the parameters chosen as $\xi_1=0.5,\eta_1=1,c_1=1$. Red line absolute value of $u$, blue line real part of $u$ and green line imaginary of $u$.}\label{f2}
\end{figure}

\begin{figure}
\centering
\includegraphics[width=3.6cm,height=3.0cm]{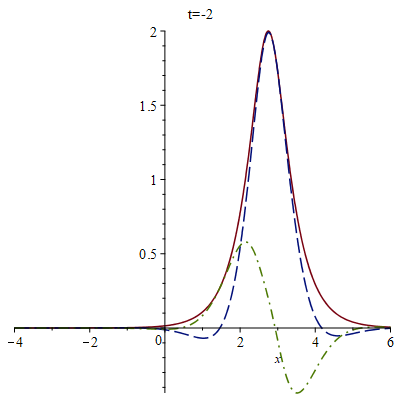}\qquad \qquad
\includegraphics[width=3.6cm,height=3.0cm]{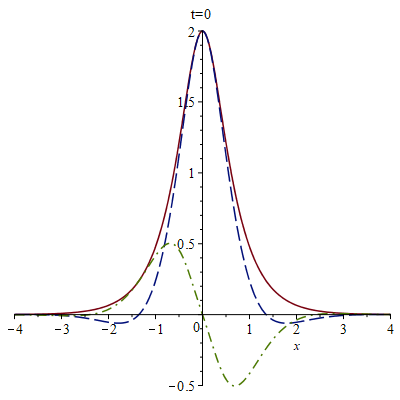}\qquad \qquad
\includegraphics[width=3.6cm,height=3.0cm]{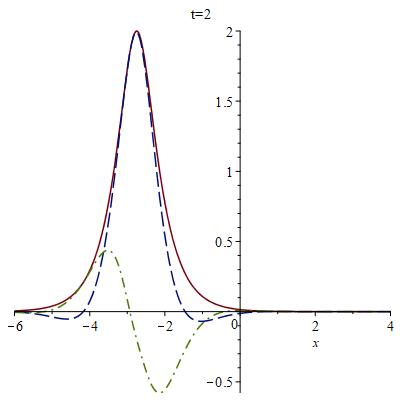}
\caption{(Color online) One-soliton solution for $|u|$, where $\xi_1=1,\eta_1=0.5,c_1=1$.}\label{f3}
\end{figure}

\begin{figure}
\centering
\includegraphics[width=3.6cm,height=3.0cm]{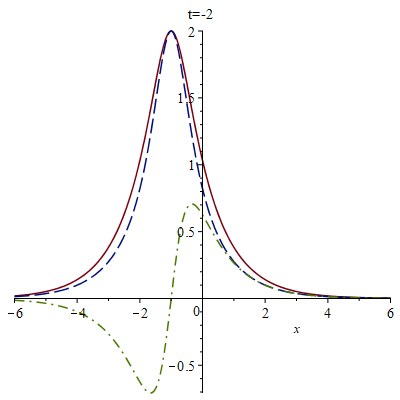}\qquad \qquad
\includegraphics[width=3.6cm,height=3.0cm]{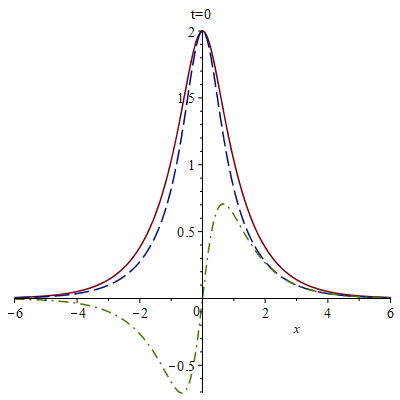}\qquad \qquad
\includegraphics[width=3.6cm,height=3.0cm]{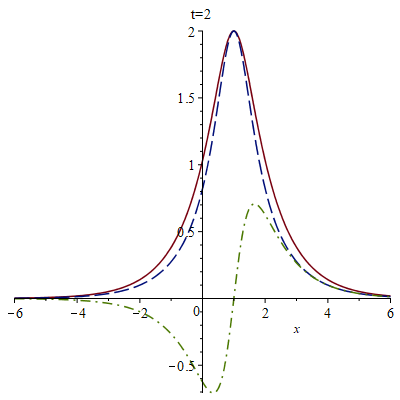}
\caption{(Color online) One-soliton solution for $|u|$, with the parameters chosen as $\xi_1=0.5,\eta_1=0.5,c_1=1$.}\label{f4}
\end{figure}

The velocity for the single soliton is $v_1 = 8\xi_1^2 \eta_1^{2}-6 (\xi_1^{2}-\eta_1^{2})^2 ,$ and its behavior occurring along the line
$$
x-v_1t+\frac{1}{4m_1}ln|c_1|=0.
$$
The amplitudes associated with $|u|^2$ are given by
 $$
 A(q)=\frac{64\xi_1^2\eta_1^2}{2|\lambda_{1}^2|+\lambda_{1}^2+\lambda_{1}^{*2}}.
 $$
 Besides, it is found that $\alpha(x)$ and $\beta(t)$ are eliminated automatically interior the calculation, so set $\alpha(x)=\beta(t)=0$
below without loss of generality. More, $\xi_1\eta_1>0$ if $\lambda_{1}\in D_{+}$, in the subregion ${\xi_1>\eta_1}$ and ${\xi_1<\eta_1}$ of $D_{+}$, the one-soliton is a  left traveling wave (see Fig.(\ref{f2}) and Fig.(\ref{f3})). On the line $\xi_1=\eta_1$, the
one-soliton is a right traveling wave (see Fig. (\ref{f4})).

Compared with the classical second-order flow GI equation(\ref{DNLS3}), for $\lambda$ belongs to $D_{+}$, soliton solutions of GI equation have three different traveling wave directions: left traveling wave, right traveling wave and stationary wave. By choosing the same parameters as Ref.\cite{NH2018AMP}, We find that for the soliton solutions of the third-order flow GI equation (\ref{hDNLS3}), there are only two kinds of traveling wave solutions, left traveling wave and right traveling wave. And the wave propagation velocity also changed, the classical GI equation velocity is: $4(\xi_1^2-\eta_1^2)$, and the velocity of high-order GI equation is: $8\xi_1^2 \eta_1^{2}-6 (\xi_1^{2}-\eta_1^{2})^2 .$ Besides, the amplitude of the soliton solution of the higher-order GI equation is also affected. Compared with the classical GI, the amplitude of the third-order flow GI equation becomes higher. That is to say, the introduction of third-order dispersion and fifth-order nonlinearity will affect the velocity, direction and amplitude of solution.

\subsection{ Two-soliton solutions}
\

When $N = 2$, the two-soliton solutions of the third-order flow GI equation can be written out explicitly as follows
\begin{equation}\label{GI2}
u(x,t)=\frac{a_1e^{\Theta_1^{'}-\Theta_{2}}+a_2e^{\Theta_1^{'}+\Theta_{2}}+a_3
e^{-\Theta_1+\Theta_{2}^{'}}+a_4e^{\Theta_{1}+\Theta_{2}^{'}}}
{b_1e^{-\Theta_1-\Theta_{2}}+b_2e^{\Theta_{1}+\Theta_{2}}
+b_3e^{\Theta_1^{'}-\Theta_{2}^{'}}+b_4e^{-\Theta_1^{'}+\Theta_{2}^{'}}
+b_5e^{\Theta_1-\Theta_{2}}+b_6e^{-\Theta_1+\Theta_{2}}},
\end{equation}
where
\begin{align}\nonumber
&\Theta_1=\theta_{1}+\theta_{1}^{*};~\Theta_1^{'}=\theta_{1}-\theta_{1}^{*};~\Theta_2=\theta_{2}+\theta_{2}^{*};~\Theta_2^{'}=\theta_{2}-\theta_{2}^{*};\\ \nonumber
&a_1=c_1\lambda_{2}(\lambda_{1}^2-\lambda_{1}^{*2})(\lambda_{1}^2-\lambda_{2}^{*2})(\lambda_{2}^{*2}-\lambda_{1}^{*2});\\ \nonumber
&a_2=c_1|c_2|^2\lambda_{2}^{*}(\lambda_{1}^2-\lambda_{1}^{*2})(\lambda_{2}^2-\lambda_{1}^{*2})(\lambda_{1}^2-\lambda_{2}^2);\\ \nonumber
&a_3=c_2\lambda_{1}(\lambda_{2}^2-\lambda_{2}^{*2})(\lambda_{2}^2-\lambda_{1}^{*2})(\lambda_{1}^{*2}-\lambda_{2}^{*2});\\ \nonumber
&a_4=|c_1|^2c_2\lambda_{1}^{*}(\lambda_{2}^2-\lambda_{2}^{*2})(\lambda_{1}^2-\lambda_{2}^{*2})(\lambda_{2}^2-\lambda_{1}^2);\\ \nonumber
&b_1=2\lambda_{1}\lambda_{2}(\lambda_{1}^2-\lambda_{2}^{2})(\lambda_{2}^{*2}-\lambda_{1}^{*2});\\ \nonumber
&b_2=2|c_1|^2|c_2|^2\lambda_{1}^{*}\lambda_{2}^{*}(\lambda_{1}^2-\lambda_{2}^{2})(\lambda_{2}^{*2}-\lambda_{1}^{*2});\\ \nonumber
&b_3=-2c_1c_2^{*}|\lambda_{2}|^2(\lambda_{1}^2-\lambda_{1}^{*2})(\lambda_{2}^{2}-\lambda_{2}^{*2});\\ \nonumber
&b_4=-2c_1^{*}c_2|\lambda_{1}|^2(\lambda_{1}^2-\lambda_{1}^{*2})(\lambda_{2}^{2}-\lambda_{2}^{*2});\\ \nonumber
&b_5=2|c_1|^{2}\lambda_{1}^{*}\lambda_{2}(\lambda_{1}^2-\lambda_{2}^{*2})(\lambda_{2}^{2}-\lambda_{1}^{*2});\\ \nonumber
&b_6=2|c_2|^2\lambda_{1}\lambda_{2}^{*}(\lambda_{1}^2-\lambda_{2}^{*2})(\lambda_{2}^{2}-\lambda_{1}^{*2}). \nonumber
\end{align}

We show the typical solution behaviors in Fig. 5 with $\lambda_1=1+0.3i, c_1=1, \lambda_2=1+0.5i, c_2=1$.
\begin{figure}
\centering
\includegraphics[width=3.8cm,height=3.6cm]{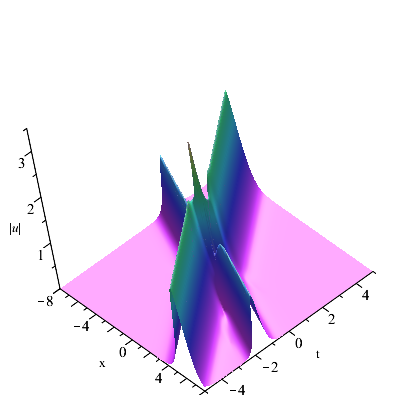}\qquad \qquad
\includegraphics[width=3.8cm,height=3.6cm]{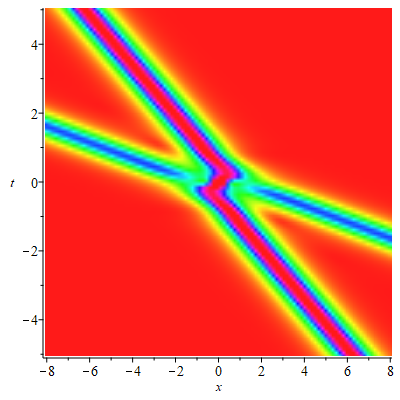}\qquad \qquad
\includegraphics[width=3.8cm,height=3.6cm]{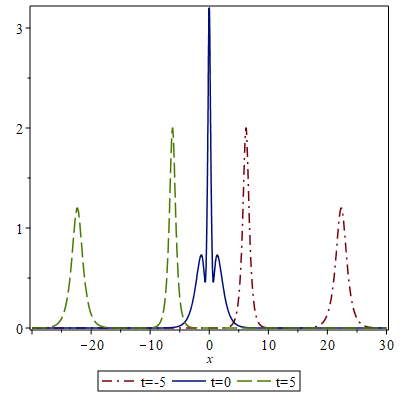}\\
$(a)$\qquad \qquad \qquad \qquad\qquad \qquad \qquad$(b)$\qquad \qquad\qquad \qquad \qquad\qquad \qquad \qquad$(c)$
\caption{(Color online) Two-soliton solution for $|u|$, (a) Three dimensional plot; (b) The density plot; (c) The plot for the 2-soliton solution evolution. where $\lambda_1=1+0.3i, c_1=1, \lambda_2=-1+0.5i, c_2=1.$}\label{f5}
\end{figure}

We see from Fig.(\ref{f5})(a) that as $t\rightarrow-\infty$, the solution consists of two single-solitons which are far apart and moving toward each other. When they collide, they interact strongly. But when $t\rightarrow\infty$, these solitons reemerge out of interactions without any change of shape and velocity, and there is no energy radiation emitted to the far field. Thus the interaction of these solitons is elastic.  Indeed, after the interaction, each soliton acquires a position shift and a phase shift. The position of each soliton is always shifted forward, as if the soliton accelerates during interactions.

To show this fact, we analyze the asymptotic states of the solution (\ref{L20}) as $t\rightarrow\pm\infty$. Without loss of generality, let us assume that $\xi_i\eta_i>0$ and $v_{1}<v_{2}$. This means that at $t\rightarrow-\infty$, soliton-1 is on the right side of soliton-2 and moves slower.  In the moving frame with velocity $v_{i}=8\xi_i^2 \eta_i^{2}-6 (\xi_i^{2}-\eta_i^{2})^2 $, note that $z_1=2 m_{1}(x-v_{1} t),z_2=2 m_{2}(x-v_{2}t)$, it yields
$$
m_2z_1-m_1z_2=2m_1m_2(v_{2}-v_{1})t.
$$
When $t \rightarrow-\infty$, $|z_1|<\infty, z_2\rightarrow+\infty$. In this case, simple calculations show that the asymptotic state of the solution (\ref{GI2}) is
$$
u(x,t)\rightarrow-2i(\lambda_{1}^2-\lambda_{1}^{*2}) \frac{c_{1}^{-} e^{\theta_{1}-\theta_{1}^{*}}}{\lambda_{1} e^{-(\theta_{1}+\theta_{1}^{*})}+\lambda_{1}^{*}|c_{1}^{-}|^{2} e^{\theta_{1}+\theta_{1}^{*}}},~~~~t\rightarrow-\infty,
$$
where $c_1^{-}=c_1\frac{(\lambda_{1}^2-\lambda_{2}^{2})}{(\lambda_{1}^2-\lambda_{2}^{*2})}$. Comparing this expression with (\ref{L21}), we see that this asymptotic solution is a single-soliton solution with peak amplitude $\frac{64\xi_1^2\eta_1^2}{2|\lambda_{1}^2|+\lambda_{1}^2+\lambda_{1}^{*2}}$ and velocity $8\xi_1^2 \eta_1^{2}-6 (\xi_1^{2}-\eta_1^{2})^2$.

 When $t \rightarrow+\infty$, $|z_1|<\infty, z_2\rightarrow-\infty$. In this case, the asymptotic state of the solution (\ref{GI2}) is
$$
u(x,t)\rightarrow-2i(\lambda_{1}^2-\lambda_{1}^{*2}) \frac{c_{1}^{+} e^{\theta_{1}-\theta_{1}^{*}}}{\lambda_{1} e^{-(\theta_{1}+\theta_{1}^{*})}+\lambda_{1}^{*}|c_{1}^{+}|^{2} e^{\theta_{1}+\theta_{1}^{*}}},~~~~t\rightarrow+\infty,
$$
where $c_1^{+}=c_1\frac{(\lambda_{1}^2-\lambda_{2}^{*2})}{(\lambda_{1}^2-\lambda_{2}^{2})}$. This is also a single-soliton solution with peak amplitude $\frac{64\xi_1^2\eta_1^2}{2|\lambda_{1}^2|+\lambda_{1}^2+\lambda_{1}^{*2}}$ and velocity $8\xi_1^2 \eta_1^{2}-6 (\xi_1^{2}-\eta_1^{2})^2$. This indicates that this soliton does not change its shape and velocity after collision. Its position and phase have shifted, however, as Fig. (\ref{f5})(b) has shown. The position shift is
$$
\Delta x_{01}=-\frac{1}{8\xi_1\eta_1}\left(\ln \left|c_{1}^{+}\right|-\ln \left|c_{1}^{-}\right|\right)=\frac{1}{4\xi_1\eta_1} \ln \left|\frac{\lambda_{1}^2-\lambda_{2}^{2}}{\lambda_{1}^2-\lambda_{2}^{*2}}\right|,
$$
and the phase shift is
$$
\Delta \sigma_{01}=\arg \left(c_{1}^{+}\right)-\arg \left(c_{1}^{-}\right)=-2 \arg \left(\frac{\lambda_{1}^2-\lambda_{2}^{2}}{\lambda_{1}^2-\lambda_{2}^{*2}}\right).
$$
Notice that $\Delta x_{01}<0 $ since $ \lambda_{k}\in D_{+}$, and thus the (slower) soliton-1 acquires a negative position shift.

Following similar calculations, we find that soliton-2 in the moving frame with velocity $8\xi_2^2 \eta_2^{2}-6 (\xi_2^{2}-\eta_2^{2})^2$, as $t \rightarrow\pm\infty$, the asymptotic solutions are both single soliton with the same peak amplitude  $8\xi_2\eta_2$, and the soliton constants $c_2^{\pm}$ before and after collision are related as
$$
c_2^{+}=c_2^{-}\frac{(\lambda_{1}^2-\lambda_{2}^{2})^2}{(\lambda_{1}^{*2}-\lambda_{2}^{2})^2}.
$$
Thus, after collision, this second soliton acquires a position shift
$$
\Delta x_{02}=-\frac{1}{8\xi_2\eta_2}\left(\ln \left|c_{2}^{+}\right|-\ln \left|c_{2}^{-}\right|\right)=-\frac{1}{4\xi_2\eta_2} \ln \left|\frac{\lambda_{1}^2-\lambda_{2}^{2}}{\lambda_{1}^{*2}-\lambda_{2}^{2}}\right|,
$$
and a phase shift
$$
\Delta \sigma_{02}=\arg \left(c_{2}^{+}\right)-\arg \left(c_{2}^{-}\right)=2 \arg \left(\frac{\lambda_{1}^2-\lambda_{2}^{2}}{\lambda_{1}^{*2}-\lambda_{2}^{2}}\right).
$$
Notice that $\Delta x_{02}>0 $, indicating that the (faster) soliton-2 acquires a positive position shift.
In addition,
$$
-\frac{\Delta x_{02}}{\Delta x_{01}}=\frac{\xi_1\eta_{1}}{\xi_2\eta_{2}},
$$
thus the amount of each soliton's position shift is inversely proportional to its amplitude.

\section{Soliton matrices for high-order zeros}

 In this section, we will consider the high-order zeros in RHP of the third-order flow GI equation. We assume $\operatorname{det}P_{+}(\lambda)$ have high-order zeros $\{\pm \lambda_{j} \}_{j=1}^{N}$, from the symmetries (\ref{2.36})
and (\ref{2.37}), we know that $\{\pm \lambda_{j}^{*}\}_{j=1}^{N}$ are high-order zeros of $\operatorname{det}P_{-}^{-1}(\lambda)$. So $\operatorname{det}P_{+}(\lambda)$ and $\operatorname{det}P_{-}^{-1}(\lambda)$
can be expanded as:
$$
\operatorname{det}P_{+}(\lambda)=s_{11}(\lambda)=\left(\lambda^{2}-\lambda_{1}^{2}\right)^{n_{1}}\left(\lambda^{2}-\lambda_{2}^{2}\right)^{n_{2}} \cdots\left(\lambda^{2}-\lambda_{N}^{2}\right)^{n_{N}} s_{0}(\lambda),
$$
$$
\operatorname{det}P_{-}^{-1}(\lambda)=\hat{s}_{11}(\lambda)=\left(\lambda^{2}-\lambda_{1}^{*2}\right)^{n_{1}}\left(\lambda^{2}-\lambda_{2}^{*2}\right)^{n_{2}} \cdots\left(\lambda^{2}-\lambda_{N}^{*2}\right)^{n_{N}} \hat{s}_{0}(\lambda),
$$
 where $s_0(\lambda)\neq 0$ for all $\lambda \in D_{+}$, and $\hat{s}_0(\lambda)\neq 0$ for all $\lambda \in D_{-}$.

 First of all, we let functions $P_{+}(\lambda)$ and $P_{-}^{-1}(\lambda)$ from above RHP have only one pair of zero of order $n_1$, i.e. $\{\lambda_{1},-\lambda_{1}\}$ and $\{\lambda_{1}^{*},-\lambda_{1}^{*}\}$. Hence, one needs to construct the dressing matrix $\Gamma(\lambda)$ whose determinant is $\frac{(\lambda^2-\lambda_{1}^2)^{n_1}}{(\lambda^2-\lambda_{1}^{*2})^{n_1}}$. For multiple zeros, its kernel vector will no longer be one. The geometric multiplicity of $\pm \lambda_i (\pm \lambda_i^{*})$ is defined as the number of the null vectors in the kernel of $\operatorname{det}P_{+}$( $\operatorname{det}P_{-}^{-1}$). It can be easily shown that the order of a zero is always greater or equal to its geometric multiplicity. It is also obvious that the geometric multiplicity of a zero is less than the matrix dimension.

Below we derive the soliton matrix $\Gamma(\lambda)$ and its inverse for an elementary high-order zero. The results are presented in the following lemma.

\noindent \textbf{Lemma 1}  \emph{(\cite{YB2019NAR},Lemma 1)
 Consider a pair of elementary high-order zeros of order $n$: $\{\lambda_1,-\lambda_1\}$ in $D_{+}$ and $\{\lambda_1^{*},-\lambda_1^{*}\}$ in $D_{-}$. Then the corresponding soliton matrix $\Gamma(\lambda)$  and its inverse can be cast in the following form
 \begin{equation}\label{5.16}
 \begin{array}{l}
\Gamma^{-1}(\lambda)=I+\left(\left|p_{1}\right\rangle, \cdots,\left|{\tilde{p}}_{n}\right\rangle\right) \mathcal{D}(\lambda)\left(\begin{array}{c}
\left\langle q_{n}\right| \\
\vdots \\
\left\langle{\tilde{q}}_{1}\right|
\end{array}\right), \\
\Gamma(\lambda)=I+\left(\left|\bar{q}_{n}\right\rangle, \cdots,\left|\bar{\tilde{q}}_{1}\right\rangle\right) \overline{\mathcal{D}}(\lambda)\left(\begin{array}{c}
\left\langle\bar{p}_{1}\right| \\
\vdots \\
\left\langle{\overline{\tilde{p}}_{n}}\right|
\end{array}\right),
\end{array}
 \end{equation}
 where the matrices $\mathcal{D}(\lambda)$ and $\overline{\mathcal{D}}(\lambda)$ are defined as
 $$
\mathcal{D}(\lambda)=\left(\begin{array}{cc}
\mathcal{K}^{+}\left(\lambda-\lambda_{1}\right) & \mathbf{0}_{n \times n} \\
\mathbf{0}_{n \times n} & \mathcal{K}^{+}\left(\lambda+\lambda_{1}\right)
\end{array}\right), \quad \overline{\mathcal{D}}(\lambda)=\left(\begin{array}{cc}
\mathcal{K}^{-}\left(\lambda-\lambda_{1}^{*}\right) & \mathbf{0}_{n \times n} \\
\mathbf{0}_{n \times n} & \mathcal{K}^{-}\left(\lambda+\lambda_{1}^{*}\right)
\end{array}\right),
$$
$\mathcal{K}^{+}(s),\mathcal{K}^{-}(s)$are upper-triangular and lower-triangular Toeplitz matrices defined as:
$$
\mathcal{K}^{+}(s)=\left(\begin{array}{cccc}
s^{-1} & s^{-2} & \cdots & s^{-n} \\
0 & \ddots & \ddots & \vdots \\
\vdots & \ddots & s^{-1} & s^{-2} \\
0 & \cdots & 0 & s^{-1}
\end{array}\right),~~~~~~\mathcal{K}^{-}(s)=\left(\begin{array}{cccc}
s^{-1} & 0 & \cdots & 0 \\
s^{-2} & s^{-1} & \ddots & \vdots \\
\vdots & \ddots & \ddots & 0 \\
s^{-n} & \cdots & s^{-2} & s^{-1}
\end{array}\right),
$$
and vectors $|p_{j}\rangle, |\tilde{p}_{j}\rangle, \langle\bar{p}_{j}|,\langle q_{j}|,|\bar{q}_{j}\rangle, |\bar{\tilde{q}}_{j}\rangle(j=1, \ldots, n)$ are independent of $\lambda$.
}

In fact, the rest of the vector parameters in (\ref{5.16}) can be derived by calculating the poles of each order in the identity $\Gamma(\lambda) \Gamma^{-1}(\lambda)=I$ at $\lambda=\lambda_{1}$ and $\lambda=-\lambda_{1}$,
$$
\Gamma\left(\lambda_{1}\right)\left(\begin{array}{c}
\left|p_{1}\right\rangle \\
\vdots \\
\left|p_{n}\right\rangle
\end{array}\right)=0, \quad \Gamma\left(-\lambda_{1}\right)\left(\begin{array}{c}
\left|{\tilde{p}}_{1}\right\rangle \\
\vdots \\
\left|{\tilde{p}}_{n}\right\rangle
\end{array}\right)=0,
$$
where
$$
\Gamma(\lambda)=\left(\begin{array}{cccc}
\Gamma(\lambda) & 0 & \cdots & 0 \\
\frac{d}{d \lambda} \Gamma(\lambda) & \Gamma(\lambda) & \ddots & \vdots \\
\vdots & \ddots & \ddots & 0 \\
\frac{1}{(n-1) !} \frac{d^{n-1}}{d \lambda^{n-1}} \Gamma(k) & \cdots & \frac{d}{d\lambda} \Gamma(\lambda) & \Gamma(\lambda)
\end{array}\right).
$$
Hence, in terms of the independent vector parameters, results (\ref{5.16}) can be formulated in a more compact form as in Ref. \cite{VSS2003SAM}, and here we just avoid these overlapped parts. Using this method, the process of solving soliton solution is very complex. In the following, we derive dressing matrix of higher-order poles via the method of  unipolar point limit. The specific results are given by the following theorem.

\noindent \textbf{Theorem 2}  \emph{
In the case of one pair of elementary high-order zero, the dressing matrix for the third-order flow GI equation can be represented as:
$$
\Gamma=\Gamma_{1}^{[n-1]} \cdots \Gamma_{1}^{[0]}, \quad \Gamma^{-1}=\Gamma_{1}^{[0]-1} \cdots \Gamma_{1}^{[n-1]-1},
$$
where
$$
\Gamma_{1}^{[j]}=I+\frac{A_{1}^{[j]}}{\lambda-\lambda_1^{*}}-\frac{\sigma_{3} A_{1}^{[j]} \sigma_{3}}{\lambda+\lambda_1^{*}}, \quad \Gamma_{1}^{[j]-1}=I+\frac{A_{1}^{\dagger[j]}}{\lambda-\lambda_{1}}-\frac{\sigma_{3} A_{1}^{\dagger[j]} \sigma_{3}}{\lambda+\lambda_{1}}
$$
$$
A_{1}^{[j]}=\frac{\lambda_{1}^2-\lambda_{1}^{*2}}{2}\left(\begin{array}{cc}
\alpha_{1}^{[j]} & 0 \\
0 & \alpha_{1}^{*[j]}
\end{array}\right)|v_{1}^{[j]}\rangle\langle v_{1}^{[j]}|,~~~(\alpha_{1}^{[j]})^{-1}=\langle v_{1}^{[j]}|\left(\begin{array}{cc}
\lambda_1^{*} & 0 \\
0 & \lambda_1
\end{array}\right)|v_{1}^{[j]}\rangle,
$$
and
$$
|v_{1}^{[j]}\rangle=\lim_{\delta \rightarrow 0} \frac{(\Gamma_{1}^{[j-1]} \cdots \Gamma_{1}^{[0]})|_{\lambda=\lambda_{1}+\delta}}{\delta^{j}}|v_{1}\rangle(\lambda_{1}+\delta),
$$
$$
\langle v_{1}^{[j]}|=\lim_{\delta \rightarrow 0}\langle v_{1}|(\lambda_{1}^{*}+\delta) \frac{(\Gamma_{1}^{[0]-1} \cdots \Gamma_{1}^{[j-1]-1} |)_{\lambda=\lambda_{1}^{*}+\delta}}{\delta^{j}}.
$$
}

Then by techniques similar to those used above, we can get
$$
u= 2i\left(\sum_{j=0}^{n-1}[B_{1}^{[j]}-\sigma_{3} B_{1}^{[j]} \sigma_{3}]_{12}\right).
$$

\begin{figure}
\centering
\includegraphics[width=3.8cm,height=3.6cm]{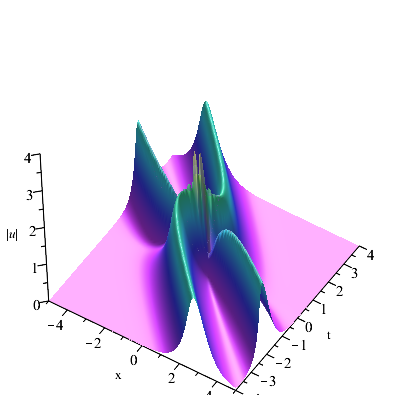}\qquad \qquad
\includegraphics[width=3.8cm,height=3.6cm]{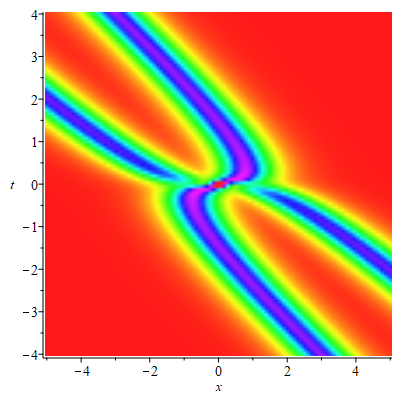}\qquad \qquad
\includegraphics[width=3.8cm,height=3.6cm]{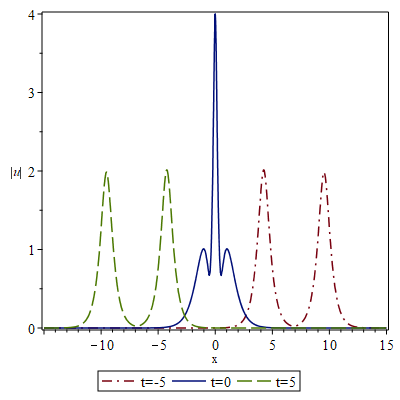}\\
$(a)$\qquad \qquad \qquad \qquad\qquad \qquad \qquad$(b)$\qquad \qquad\qquad \qquad \qquad\qquad \qquad \qquad$(c)$
\caption{(Color online) high-order 1-soliton solution for $|u|$.(a)  Three dimensional plot; (b) The density plot; (c) The plot for the high-order 1-soliton solution evolution. where $\lambda=1+0.5i, c_1=1.$}\label{f6}
\end{figure}

As before,  the above formulas also could be rewritten with the determinant form
\begin{equation}\label{Hs}
u=-4i\frac{det\tilde{F}}{det\tilde{M}},
\end{equation}
where
$$
\tilde{F}=\left(\begin{array}{ccccc}
\tilde{M}_{11} & \tilde{M}_{12} & \cdots & \tilde{M}_{1 n} & |v_{1}\rangle^{[0]}_1 \\
\tilde{M}_{21} & \tilde{M}_{22} & \cdots & \tilde{M}_{2 n} & |v_{1}\rangle^{[1]}_1 \\
\vdots & \vdots & \ddots & \vdots & \vdots \\
\tilde{M}_{{n}1} & \tilde{M}_{{n} 2} & \cdots & \tilde{M}_{nn} & |v_{1}\rangle^{[n-1]}_1  \\
\langle v_{1}|^{[0]}_2 & \langle v_{1}|^{[1]}_2 & \cdots & \langle v_{1}|^{[n-1]}_2 & 0
\end{array}\right),
$$
and
$$
\tilde{M}_{k l}=\frac{1}{(k-1) !(l-1) !} \frac{\partial^{k+l-2}}{\partial \lambda^{*k-1} \partial \lambda^{l-1}}\frac{\langle v_{1}\mid v_{1}\rangle}{\lambda-\lambda^{*}}-\frac{\left\langle v_{1}\left|\sigma_{3}\right| v_{1}\right\rangle}{\lambda+\lambda^{*}}|_{{\lambda=\lambda_1}, \lambda^{*}=\lambda_1^{*}}.
$$
Where $|v_{1}\rangle^{[j]}, \langle v_{1}|^{[j]}$ can be written as follows
\begin{equation}
|v_{1}\rangle^{[j]}=\frac{1}{(j)!}\frac{\partial^{j}}{\partial(\lambda)^{j}}|v_{1}\rangle|_{\lambda=\lambda_{1}},~~~\langle v_{1}|^{[j]}=\frac{1}{(j)!}\frac{\partial^{j}}{\partial(\lambda)^{j}}\langle v_{1}|_{\lambda=\lambda_{1}^{*}}.
\end{equation}
Hence, formula (\ref{Hs}) leads to the  elementary high-order zeros solution formula. When $N=1$, it corresponds to a single soliton solution, and when $N\geq2$, it corresponds to  higher-order soliton. Notice that the general expression of the high-order soliton solution of Eq. (\ref{Hs}) is very complicated and is not given explicitly. However, with the aid of computer softwares such as Maple and Matlab, one can easily get the corresponding double-pole solution for different parameters by using Eq. (\ref{Hs}). Explicitly,  taking $N = 2$, in (\ref{Hs}) by choosing appropriate parameters   considering the simplest higher-order 1-soliton solution case, which is plotted in Fig.(\ref{f6}).

\section{Conclusion and discussion}

In summary, the GI hierarchy is derived by using recursive operator. The recursive operator here contains two operators, which is more complex than the form of AKNS hierarchy and KN hierarchy. The main reason is that the derivative of the main diagonal of $M$ to $t$ is not 0, but a function related to the potential functions $u$ and $v$, which leads to the complex expression of $A$. Then the inverse scattering method has been applied to the third-order flow GI equation and by considering the associated RHP, we successfully give a simple representation for the N-soliton in the determinant form. Owing to the symmetry properties of Jost solution and scattering data, the corresponding zeros in the RHP for higher-order GI equation appear in pairs, which is the same as the $3\times3$ Sasa-Satsuma equation\cite{YB2019NAR}. Later, taking single-soliton solution and two-soliton solutions as examples, the long-time behavior of the solution is studied. Compared with the classical GI direction of the second-order flow, it is found that the motion direction and wave height of the soliton solution are affected by the third-order dispersion and the fifth-order nonlinearity. These analysis results have important reference value for the study of GI hierarchy or other nonlinear integrable dynamic systems of higher order flow equations, and provide a theoretical basis for possible experimental research and application. Finally, the corresponding higher-order soliton solution matrix is derived by analyzing the limiting behavior of spectral parameters.

In recent years, there are many achievements in the study of the classical second-order flow GI equation with non-zero boundary conditions \cite{PWQ2021AX,ZZC2021AX,ZGQ2021AX}. In this paper, we only consider the simple zeros and a pair of elementary higher-order zeros of the third-order flow GI equation with vanishing boundary conditions. Whether the behavior of soliton solutions with non-zero boundary and more multiplicity will have more abundant forms and long time behavior can be studied in the future.


\begin{thebibliography}{99}
\bibitem{Das1989Singapore}
Das A. Integrable models. World Scientific. Singapore 1989.
 \bibitem{JRS1977PRSL}
 Johnson RS. On the modulation of water waves in the neighbourhood of $kh\approx1.363$. Proc.Roy.Soc.London Ser.A 1977;357:131-141.
\bibitem{KY1985JSP} Kodama Y. Optical solitons in a monomode fiber. J. Statist. Phys. 1985;39:597-614.
\bibitem{GI1983BJP}Gerdjikov VS, Ivanov MI. A quadratic pencil of general type and nonlinear evolution equations. II. Hierarchies of
Hamiltonian structures. Bulg.J. Phys. 1983;10:130-143.
\bibitem{Kaup1978JMP}Kaup DJ, Newell AC.  An exact solution for a derivative nonlinear Schr\"{o}dinger equation. J. Math. Phys. 1978;19:798-801.
\bibitem{CLL1979PS}Chen HH, Lee YC, Liu CS. Integrability of nonlinear Hamiltonian systems by inverse scattering method. Phys.Scr.1979;20:490-492.
\bibitem{Fan2000JMP}Fan EG. Integrable evolution systems based on Gerdjikov-Ivanov equations, bi-Hamiltonian structure, finite-dimensional integrable systems and N-fold Darboux transformation. J. Math. Phys. 2000;41:7769-7782.
 \bibitem{Fan2000JPA}Fan EG. Darboux transformation and soliton-like solutions for the Gerdjikov-Ivanov equation. J. Phys. A: Math. Theor. 2000;33:6925-6933.
\bibitem{EM1989PS}Mj{\o}lhus E. Nonlinear Alfv\'{e}n waves and the DNLS equation: oblique aspects. Phys. Scr.1989;40: 227-237.
\bibitem{AGP2007AP}Agrawal GP. Nonlinear fiber optics. Nonlinear Science at the Dawn of the 21st Century. 2000;542:195-211.
 \bibitem{EG2001JPA}Fan EG. Integrable systems of derivative nonlinear Schr\"{o}dinger type and their multi-Hamiltonian structure. J. Phys. A: Math. Gen. 2001;34:513-519.
 \bibitem{EG2004CSF}Dai HH, Fan EG. Variable separation and algebro-geometric solutions of the Gerdjikov-Ivanov equation. Chaos, Solitons Fractals. 2004;22:93-101.
 \bibitem{SXH2012JMP}Xu SW, He JS. The rogue wave and breather solution of the Gerdjikov-Ivanov equation. J. Math.
Phys. 2012;53:063507.
\bibitem{YJK2010SIAM} Yang JK. Nonlinear Waves in Integrable and Nonintegrable Systems. Philadelphia: Soc. Indus. Appl. Math. 2010.
\bibitem{PD1993AM} Deift P, Zhou X. A steepest descent method for oscillatory Riemann-Hilbert problems. Annals of Mathematics, 1993;137:295-368.
\bibitem{HBB2020AR}  Hua BB, Zhang, L,Zhang N. On the Riemann-Hilbert problem for the Chen-Lee-Liu derivative nonlinear Schr\"{o}dinger equation. arXiv:2004.07608v1.
\bibitem{ZN2018AM} Zhang N, XIA TC, Fan EG. A Riemann-Hilbert Approach to the Chen-Lee-Liu Equation on the Half Line. Acta Mathematicae Applicatae Sinica, English Series.2018;34;493-515.
\bibitem{MWX2019NARWA} Ma WX. Application of the Riemann-Hilbert approach to the multicomponent AKNS integrable hierarchies. Nonlinear
Anal. Real World Appl. 2019;47:1-17.
\bibitem{TSZT2018PAMS}Tian SF, Zhang TT. Long-time asymptotic behavior for the Gerdjikov-Ivanov type of derivative
nonlinear Schr\"{o}dinger equation with time-periodic boundary condition. Proc. Amer. Math.Soc. 2018;146:1713-1729.
 \bibitem{XJ2013MPAG} Xu J, Fan EG and Chen Y. Long-time asymptotic for the derivative nonlinear Schr\"{o}dinger equation with step-like initial value. Math. Phys. Anal. Geom. 2013;16:253-288.
\bibitem{GB2012JMP} Guo BL, Ling LM. Riemann-Hilbert approach and N-soliton formula for coupled derivative Schr\"{o}dinger equation. J. Math.
Phys. 2012;53:073506.
 \bibitem{CS1967PRL}Gardner CS, Greene JM, Kruskal MD and Miura RM. Method for solving the Kortcmeg-de Vries equation. Phys. Rev. Lett. 1967;19:1095-1097.
\bibitem{AS1976MP}Shabat AB. One dimensional perturbations of a differential operator and the inverse scattering problem. Problems in Mechanics and Mathematical Physics.1976: 279-296, Nauka, Moscow .
\bibitem{MA1974MP}Ablowitz MJ, Kaup DJ, Newell AC, Segur H. The inverse scattering transform-ourier analysis for nonlinear problems. Stud. Appl. Math. 1974;53:249-315.
\bibitem{PB1984PAM} Beals R, Coifman RR. Scattering and inverse scattering for first order systems. Comm. Pure Appl. Math. 1984; 37:39-90.
\bibitem{LG1994OL}Gagnon L, Sti\'{e}venart N. N-soliton interaction in optical fibers: The multiple-pole case. Opt. Lett. 1994;19:619-621.
 \bibitem{VSS2003SAM}Shchesnovich VS, Yang JK. Higher-Order solitons in the N-wave system. Stud. Appl. Math. 2003;110:297-332.
\bibitem{ZYS2017JMP} Zhang YS. Riemann-Hilbert method and N-soliton for two-component Gerdjikov-Ivanov equation. Journal of Nonlinear Mathematical Physics.2017;24:210-223.
\bibitem{YB2019NAR} Yang B, Chen Y. High-order soliton matrices for Sasa-Satsuma equation via local Riemann-Hilbert problem, Nonlinear Anal-Real. 2019;45:918-941.
 \bibitem{SN1984CB}Novikov S, Manakov SV, Pitaevskii LP, Zakharov VE. Theory of solitons: the inverse scattering
method (New York and London, Consultants Bureau, 1984).
 \bibitem{JH2018JNMP}Hu J, Xu J, Yu GF. Riemann-Hilbert approach and N-soliton formula for a higher-order Chen-Lee-Liu equation. Journal of Nonlinear Mathematical Physics. 2018;25:633-649.
 \bibitem{NH2018AMP}Nie H, Zhu JY, Geng XG. Trace formula and new form of N-soliton to the Gerdjikov-Ivanov equation. Anal.Math.Phys. 2018; 8:415-426.
    \bibitem{PWQ2021AX} Peng WQ, Chen Y. Double poles soliton solutions for the Gerdjikov-Ivanov type of derivative nonlinear Schr\"{o}dinger equation with zero/nonzero boundary conditions. arXiv:2104.12073.
  \bibitem{ZZC2021AX} Zhang ZC, Fan EG. Inverse scattering transform and multiple high-order pole solutions for the Gerdjikov-Ivanov equation under the zero/nonzero background.arXiv:2012.13654v1.
  \bibitem{ZGQ2021AX} Zhang GQ, Yan ZY. The Derivative Nonlinear Schr\"{o}dinger Equation with Zero/Nonzero Boundary Conditions: Inverse Scattering Transforms and N-Double-Pole Solutions. J. Nonlinear Sci.2020;30:3089-3127.


\end{thebibliography}
\end{document}